\documentclass[a4paper,12pt,notitlepage]{revtex4-1}
\usepackage{graphicx}
\usepackage{amsmath,amssymb}

\usepackage[colorlinks=true]{hyperref}
\hypersetup{
    urlcolor=[rgb]{0.0, 0.0, 0.4},
    citecolor=[rgb]{0.0, 0.4, 0.0},
    linkcolor=[rgb]{0.0, 0.2, 0.4}
}

\newcommand{{\E}}{{E_0}}
\newcommand{{\xm}}{x_{\text{max}}}
\newcommand{{\lnA}}{{\left<\ln A\right>}}
\newcommand{{\degr}}{{^{\circ}}}
\newcommand{{\SVert}}{{S_{600}(0\degr)}}
\newcommand{{\STheta}}{{S_{600}(\theta)}}
\newcommand{{\depth}}{{g/cm$^2$}}
\newcommand{{\rM}}{{r_{\text{M}}}}
\newcommand{{\rhos}}{{\rho_s}}
\newcommand{{\latG}}{{b_{\rm G}}}
\newcommand{{\lonG}}{{l_{\rm G}}}

\begin{document}

\title{A beam of ultrahigh energy particles of the same magnetic rigidity}

\author{G.\,F.~Krymsky}
\email{krymsky@ikfia.ysn.ru}
\author{I.\,Ye.~Sleptsov}
\email{sleptsov@ikfia.ysn.ru}
\author{M.\,I.~Pravdin}
\email{m.i.pravdin@ikfia.ysn.ru}
\author{A.\,D.~Krasilnikov}
\email{a.d.kras@ikfia.ysn.ru}

\affiliation{Yu.\,G.~Shafer Institute of Cosmophysical Research and Aeronomy SB
RAS\\
677980 Lenin Ave. 31, Yakutsk, Russia}

\begin{abstract}
    Two EAS arrays during a day have recorded 3 particles with energies above
    30 EeV arriving from the same sky region. Two events were registered by the
    Yakutsk array and one - by the Telescope Array. Two Yakutsk events were
    estimated to be same energy and the Telescope Array shower's energy was
    almost 2 times higher. This indicates the same magnetic rigidity of all
    three particles, if the charge differs by a factor of 2. The relatively
    short sequence of all three events and their "monochromaticity" in rigidity
    can be due to the magnetic separation of particles in the acceleration
    process and propagation of the beam.
\end{abstract}

\maketitle

Earlier it was reported that a short-lived cosmic rays (CRs) particle beam with
energy above $3 \times 10^{19}$~eV was detected arriving from a compact sky
region~\cite{bib:1}. Two particles were detected at the Yakutsk EAS Array
(YEASa)~\cite{bib:2, bib:3}, one~--- by the Telescope Array (TA)~\cite{bib:4,
bib:5}.  Chance probability of random appearance of such triplet is $2.6 \times
10^{-6}$.  Parameters of registered showers are listed in Table~\ref{t:1}. The
probable mechanism for generation of such beams essentially is interaction of
CRs with relativistic shock~\cite{bib:6, bib:7}. If a shock carries a strong
magnetic field it can trap particles of ultra-high energy CRs (UHECRs) and
re-emit them. This process results in a rise of the particle flux intensity
along the axis of generated beam by magnitude proportional to the shock's
Lorentz factor ($\gamma$) to the power of a value greater than 8~\cite{bib:1}.

\begin{table}[htb]
    \caption{Parameters of shower events constituting a beam. The $E_{2003}$
        column cites energy estimation according to the formula~(\ref{eq:1}).
        Zenith angle ($\theta$), right ascension ($\alpha$), declination
        ($\delta$), galactic latitude ($\latG$) and galactic longitude
    ($\lonG$) are given in degrees.}
    \label{t:1}
    \vspace{5mm}
    \centering
    \begin{tabular}
        {
            p{0.1\textwidth}
            p{0.15\textwidth}
            p{0.12\textwidth}
            p{0.08\textwidth}
            *{5}{p{0.07\textwidth}}
        }
        \hline
        \hline
        Array & Date & Time & $E_{2003}$ & $\theta$ & $\alpha$ & $\delta$ &
        $\latG$ & $\lonG$ \\
              &      & UTC  & EeV &  & & & & \\
        \hline
        Yakutsk & Jan 21, 2009 & 23:40:35 & 36.3 & 35.5 & 356.3 & 65.8 & 4.0 &
        116.5 \\
        Yakutsk & Jan 22, 2009 & 10:51:52 & 35.5 & 42.7 & 333.3 & 62.3 & 4.9 &
        105.9 \\
        TA & Jan 22, 2009 & 22:54:22 & 57.9 & 31.3 & 311.2 & 51.1 & 5.1 & 89.5
        \\
        \hline
        \hline
    \end{tabular}
\end{table}

It's worth noting that energies of two events recorded by YEASa are virtually
identical and energy of TA's shower is two times higher. The energy $\E$ of
Yakutsk events (given in Table~\ref{t:1}) was estimated according to a relation
presented in 2003 which was obtained with the use of calorimetric
method~\cite{bib:8}:

\begin{equation}
    E_{2003} = (4.6 \pm 1.2) \times 10^{17} \cdot \SVert^{0.98 \pm
    0.03}\text{,}
    \label{eq:1}
\end{equation}
where $\SVert$ is experimentally determined parameter~--- particle density at
600~m from shower axis. Formula (\ref{eq:1}) was recently refined~\cite{bib:9,
bib:10} and the following relation was obtained:

\begin{equation}
    E_{2014} = (3.6 \pm 1.0) \times 10^{17} \cdot \SVert^{1.02 \pm
    0.03}\text{.}
    \label{eq:2}
\end{equation}
Formula (\ref{eq:2}) does not significantly change the final $\E$ value,
since decrease of a constant is compensated by increased power-law dependency
of $S_{600}$ parameter.

\begin{figure}[htb]
    \centering
    \includegraphics[width=0.82\textwidth]{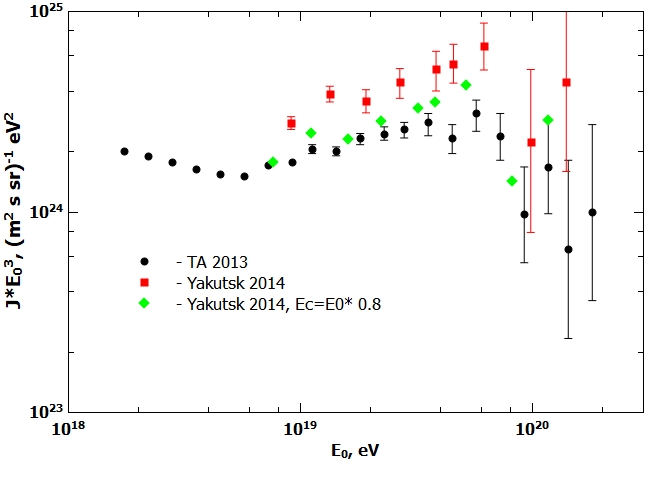}
    \caption{A comparison of CR energy spectrum according to the data from TA
        (black circles) and YEASa (red squares). Green diamonds represent the
        Yakutsk spectrum with estimated energy reduced by 20\,\%.}
    \label{fig:1}
\end{figure}

Systematic errors of energy reconstruction are presented in both experiments.
In Yakutsk it amounts to 25\,\%, at TA~--- it is 20\,\%~\cite{bib:11}, hence
there might be a systematic difference between energy scales of two
arrays; possible hint to this is the difference in CR energy spectrum
intensities in combination with close reproduction of the spectrum
shape~\cite{bib:9, bib:10}.  On Fig.~\ref{fig:1} are shown CR energy spectra of
both experiments in energy range above $10^{19}$~--- according to
TA~\cite{bib:11} and YEASa.  Yakutsk points (red squares) lie higher than those
of the TA (black circles).  If one lowers the estimated energy of Yakutsk
events by factor 0.8, then results of both experiments would agree with each
other (green diamonds). This shows that this lessening might improve the
agreement between energy scales of YEASa and TA.  That said, such a correction
wouldn't exceed one sigma in expression (\ref{eq:2}).

Ratios between TA shower energy and different energy estimations of YEASa
events are listed in Table~\ref{t:2}. In first two columns the energy is
estimated according to expression (\ref{eq:1}), as in Table~\ref{t:1}; in next
two columns a refined formula (\ref{eq:2}) was used. In next two columns energy
estimation according to (\ref{eq:2}) is multiplied by ratio 0.8. With respect
to this correction it is more probable that the energy of shower registered by
TA is approximately 2 times higher than energy of both YEASa's events. For
extreme energies such condition is satisfied when magnetic rigidity of all
three particles is the same and if electric charge of the Telescope Array
particle is 2 times higher than those registered by Yakutsk experiment. Hence
we should have two protons and alpha particle.

\begin{table}[htb]
    \caption{Ratios between energy of the TA particle and energies of
    Yakutsk array events for different estimations.}
    \label{t:2}
    \vspace{5mm}
    \centering
     \begin{tabular}
        {
            *{2}{p{0.08\textwidth}p{0.14\textwidth}}
            p{0.20\textwidth}p{0.16\textwidth}
        }
        \hline
        \hline
        $E_{2003}$ & $E_{\text{TA}} / E_{2003}$ & $E_{2014}$ & $E_{\text{TA}} /
        E_{2014}$ & $E_{\rm C} = E_{2014} \times 0.8$ & $E_{\text{TA}} / E_{\rm
        C}$ \\ 
        EeV        &            & EeV &      & EeV &  \\
        \hline
        36.3 & 1.60 & 34.2 & 1.69 & 27.4 & 2.12 \\
        35.5 & 1.63 & 33.4 & 1.73 & 26.7 & 2.17 \\
        \hline
        \hline
    \end{tabular}
\end{table}

\begin{figure}[htb]
    \centering
    \includegraphics[width=0.85\textwidth]{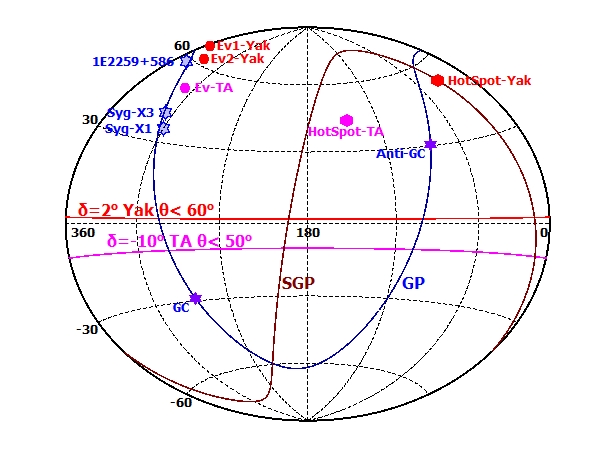}
    \caption{Location of showers from assumed beam on celestial sphere in
        equatorial coordinates. Red circles denote Yakutsk events, purple~---
        TA. Stars indicate X-ray sources with adjacent EAS arrival directions.
        Blue line represents Galaxy plane (GP), brown (SGP)~--- Super Galaxy
        plane.  HotSpot-TA is the center of region with increased particle flux
        with energy above 57~EeV according to the Telescope Array~\cite{bib:5},
        HotSpot-Yak~--- center of region with increased flux according to
        the Yakutsk array in energy range 10-30~EeV~\cite{bib:16}.
    }
    \label{fig:2}
\end{figure}

These three showers hint at existence of short-lived beams of UHECRs (with
energy
above $10^{19}$~eV) with virtually similar magnetic rigidity arriving from a
small sky region. Such beams probably are generated during rapid explosive
precesses with huge energy output. A relatively rapid succession of all three
events near Earth and their ``monochromaticism'' in magnetic rigidity are most
likely conditioned by magnetic separation of particles during propagation
of a beam. If one limits the region of the sky sphere with 15$\degr$ radius
containing all three events and assumes that particle emission within this
region is uniform; and if a burst lasts for exactly 1 day, then according to
YEASa the UHECRs the luminosity of this region increases by factor $(0.8 + 1.0 /
-0.5) \times 10^4$ compared to the background value. The Telescope Array data
published earlier~\cite{bib:5} include events with energy above 57~EeV. It is
possible that this experiment have registered events of lower energy that could
be related to the particle beam of our interest. Expansion of TA's events
selection towards lower energies and testing the set for presence of such
additional events would be instrumental in obtaining a definitive evidence that
assumed particle beam is real after all.

Locations of showers on celestial sphere in equatorial coordinate system are
shown on Figure~\ref{fig:2}. Events recorded by Yakutsk experiment are
represented with red circles, TA's particle~--- with purple circle. Stars
denote X-ray sources with adjacent EAS arrival directions. EAS events lie very
tight to the 1E2259+586 X-ray pulsar. But also they are quite close to a more
interesting source~--- Cygnus X-3 (Cyg X3); this source radiates in a wide
range~--- from radio waves to ultra-high energy photons~\cite{bib:12, bib:13}.
In early 1980's fluxes of gamma-photons with energy $10^{15}-10^{16}$~eV were
detected from Cyg~X3~\cite{bib:14, bib:15}~--- an indication of UHECRs
production.

\acknowledgements

This work is supported by the program of Presidium of RAS ``High energy physics                                                                       
and neutrino astronomy'' and by Russian Foundation for Basic Research (grant                                                                          
16--29--13019~ofi-m).

The authors would like to thank the staff of the Yakutsk EAS Array who've been
providing its operation capability for 40~years.

\bibliography{pbeam}

\begin{thebibliography}{16}%
\makeatletter
\providecommand \@ifxundefined [1]{%
 \@ifx{#1\undefined}
}%
\providecommand \@ifnum [1]{%
 \ifnum #1\expandafter \@firstoftwo
 \else \expandafter \@secondoftwo
 \fi
}%
\providecommand \@ifx [1]{%
 \ifx #1\expandafter \@firstoftwo
 \else \expandafter \@secondoftwo
 \fi
}%
\providecommand \natexlab [1]{#1}%
\providecommand \enquote  [1]{``#1''}%
\providecommand \bibnamefont  [1]{#1}%
\providecommand \bibfnamefont [1]{#1}%
\providecommand \citenamefont [1]{#1}%
\providecommand \href@noop [0]{\@secondoftwo}%
\providecommand \href [0]{\begingroup \@sanitize@url \@href}%
\providecommand \@href[1]{\@@startlink{#1}\@@href}%
\providecommand \@@href[1]{\endgroup#1\@@endlink}%
\providecommand \@sanitize@url [0]{\catcode `\\12\catcode `\$12\catcode
  `\&12\catcode `\#12\catcode `\^12\catcode `\_12\catcode `\%12\relax}%
\providecommand \@@startlink[1]{}%
\providecommand \@@endlink[0]{}%
\providecommand \url  [0]{\begingroup\@sanitize@url \@url }%
\providecommand \@url [1]{\endgroup\@href {#1}{\urlprefix }}%
\providecommand \urlprefix  [0]{URL }%
\providecommand \Eprint [0]{\href }%
\providecommand \doibase [0]{http://dx.doi.org/}%
\providecommand \selectlanguage [0]{\@gobble}%
\providecommand \bibinfo  [0]{\@secondoftwo}%
\providecommand \bibfield  [0]{\@secondoftwo}%
\providecommand \translation [1]{[#1]}%
\providecommand \BibitemOpen [0]{}%
\providecommand \bibitemStop [0]{}%
\providecommand \bibitemNoStop [0]{.\EOS\space}%
\providecommand \EOS [0]{\spacefactor3000\relax}%
\providecommand \BibitemShut  [1]{\csname bibitem#1\endcsname}%
\let\auto@bib@innerbib\@empty
\bibitem [{\citenamefont {Krymsky}\ \emph {et~al.}(2017)\citenamefont
  {Krymsky}, \citenamefont {Pravdin},\ and\ \citenamefont {Sleptsov}}]{bib:1}%
  \BibitemOpen
  \bibfield  {author} {\bibinfo {author} {\bibfnamefont {G.~F.}\ \bibnamefont
  {Krymsky}}, \bibinfo {author} {\bibfnamefont {M.~I.}\ \bibnamefont
  {Pravdin}}, \ and\ \bibinfo {author} {\bibfnamefont {I.~E.}\ \bibnamefont
  {Sleptsov}},\ }\href {\doibase 10.1134/S1063773717100048} {\bibfield
  {journal} {\bibinfo  {journal} {{Astron. Lett.}}\ }\textbf {\bibinfo {volume}
  {43}},\ \bibinfo {pages} {703} (\bibinfo {year} {2017})}\BibitemShut
  {NoStop}%
\bibitem [{\citenamefont {Artamonov}\ \emph {et~al.}(1994)\citenamefont
  {Artamonov}, \citenamefont {Afanasiev}, \citenamefont {Glushkov} \emph
  {et~al.}}]{bib:2}%
  \BibitemOpen
  \bibfield  {author} {\bibinfo {author} {\bibfnamefont {V.~P.}\ \bibnamefont
  {Artamonov}}, \bibinfo {author} {\bibfnamefont {B.~N.}\ \bibnamefont
  {Afanasiev}}, \bibinfo {author} {\bibfnamefont {A.~V.}\ \bibnamefont
  {Glushkov}},  \emph {et~al.},\ }\href@noop {} {\bibfield  {journal} {\bibinfo
   {journal} {{Bull. Russ. Acad. Sci. Phys. Trans.}}\ }\textbf {\bibinfo
  {volume} {58}},\ \bibinfo {pages} {2026} (\bibinfo {year}
  {1994})}\BibitemShut {NoStop}%
\bibitem [{\citenamefont {Ivanov}\ \emph {et~al.}(2010)\citenamefont {Ivanov},
  \citenamefont {Knurenko}, \citenamefont {Pravdin},\ and\ \citenamefont
  {Sleptsov}}]{bib:3}%
  \BibitemOpen
  \bibfield  {author} {\bibinfo {author} {\bibfnamefont {A.~A.}\ \bibnamefont
  {Ivanov}}, \bibinfo {author} {\bibfnamefont {S.~P.}\ \bibnamefont
  {Knurenko}}, \bibinfo {author} {\bibfnamefont {M.~I.}\ \bibnamefont
  {Pravdin}}, \ and\ \bibinfo {author} {\bibfnamefont {I.~E.}\ \bibnamefont
  {Sleptsov}},\ }\href {\doibase 10.3103/S0027134910040089} {\bibfield
  {journal} {\bibinfo  {journal} {{Moscow University Phys. Bull.}}\ }\textbf
  {\bibinfo {volume} {65}},\ \bibinfo {pages} {292} (\bibinfo {year}
  {2010})}\BibitemShut {NoStop}%
\bibitem [{\citenamefont {Abu-Zayyad}\ \emph {et~al.}()\citenamefont
  {Abu-Zayyad}, \citenamefont {Aida}, \citenamefont {Allen} \emph
  {et~al.}}]{bib:4}%
  \BibitemOpen
  \bibfield  {author} {\bibinfo {author} {\bibfnamefont {T.}~\bibnamefont
  {Abu-Zayyad}}, \bibinfo {author} {\bibfnamefont {R.}~\bibnamefont {Aida}},
  \bibinfo {author} {\bibfnamefont {M.}~\bibnamefont {Allen}},  \emph
  {et~al.},\ }\href {\doibase 10.1016/j.nima.2012.05.079} {\bibfield  {journal}
  {\bibinfo  {journal} {{Nucl. Instr. Meth. A}}\ }\textbf {\bibinfo {volume}
  {689}},\ \bibinfo {pages} {87}}\BibitemShut {NoStop}%
\bibitem [{\citenamefont {Abbasi}\ \emph {et~al.}(2014)\citenamefont {Abbasi},
  \citenamefont {Abe}, \citenamefont {Abu-Zayyad} \emph {et~al.}}]{bib:5}%
  \BibitemOpen
  \bibfield  {author} {\bibinfo {author} {\bibfnamefont {R.~U.}\ \bibnamefont
  {Abbasi}}, \bibinfo {author} {\bibfnamefont {M.}~\bibnamefont {Abe}},
  \bibinfo {author} {\bibfnamefont {T.}~\bibnamefont {Abu-Zayyad}},  \emph
  {et~al.},\ }\href
  {{http://iopscience.iop.org/article/10.1088/2041-8205/790/2/L21/}} {\bibfield
   {journal} {\bibinfo  {journal} {{Astrophys. J. Lett.}}\ }\textbf {\bibinfo
  {volume} {790}},\ \bibinfo {pages} {{L21}} (\bibinfo {year}
  {2014})}\BibitemShut {NoStop}%
\bibitem [{\citenamefont {Achterberg}\ \emph {et~al.}(2001)\citenamefont
  {Achterberg}, \citenamefont {Gallant}, \citenamefont {Kirk},\ and\
  \citenamefont {Guthmann}}]{bib:6}%
  \BibitemOpen
  \bibfield  {author} {\bibinfo {author} {\bibfnamefont {A.}~\bibnamefont
  {Achterberg}}, \bibinfo {author} {\bibfnamefont {Y.~A.}\ \bibnamefont
  {Gallant}}, \bibinfo {author} {\bibfnamefont {J.~K.}\ \bibnamefont {Kirk}}, \
  and\ \bibinfo {author} {\bibfnamefont {A.~W.}\ \bibnamefont {Guthmann}},\
  }\href {\doibase 10.1046/j.1365-8711.2001.04851.x} {\bibfield  {journal}
  {\bibinfo  {journal} {{MNRAS}}\ }\textbf {\bibinfo {volume} {328}},\ \bibinfo
  {pages} {393} (\bibinfo {year} {2001})},\ \Eprint
  {http://arxiv.org/abs/astro-ph/0107530} {{arXiv}:astro-ph/0107530}
  \BibitemShut {NoStop}%
\bibitem [{\citenamefont {Caprioli}(2015)}]{bib:7}%
  \BibitemOpen
  \bibfield  {author} {\bibinfo {author} {\bibfnamefont {D.}~\bibnamefont
  {Caprioli}},\ }\href
  {{http://iopscience.iop.org/article/10.1088/2041-8205/811/2/L38/}} {\bibfield
   {journal} {\bibinfo  {journal} {{Astrophys. J. Lett.}}\ }\textbf {\bibinfo
  {volume} {811}},\ \bibinfo {pages} {{L38}} (\bibinfo {year}
  {2015})}\BibitemShut {NoStop}%
\bibitem [{\citenamefont {Glushkov}\ \emph {et~al.}(2003)\citenamefont
  {Glushkov}, \citenamefont {Ivanov}, \citenamefont {Knurenko}, \citenamefont
  {Kolosov}, \citenamefont {Makarov} \emph {et~al.}}]{bib:8}%
  \BibitemOpen
  \bibfield  {author} {\bibinfo {author} {\bibfnamefont {A.~V.}\ \bibnamefont
  {Glushkov}}, \bibinfo {author} {\bibfnamefont {A.~A.}\ \bibnamefont
  {Ivanov}}, \bibinfo {author} {\bibfnamefont {S.~P.}\ \bibnamefont
  {Knurenko}}, \bibinfo {author} {\bibfnamefont {V.~A.}\ \bibnamefont
  {Kolosov}}, \bibinfo {author} {\bibfnamefont {I.~T.}\ \bibnamefont
  {Makarov}},  \emph {et~al.},\ }in\ \href
  {http://www-rccn.icrr.u-tokyo.ac.jp/icrc2003/PROCEEDINGS/PDF/99.pdf} {\emph
  {\bibinfo {booktitle} {{Proc. of the 28th ICRC, Tsukuba, Japan}}}}\ (\bibinfo
  {year} {2003})\ pp.\ \bibinfo {pages} {393--396}\BibitemShut {NoStop}%
\bibitem [{\citenamefont {Glushkov}\ \emph {et~al.}(2014)\citenamefont
  {Glushkov}, \citenamefont {Pravdin},\ and\ \citenamefont {Saburov}}]{bib:9}%
  \BibitemOpen
  \bibfield  {author} {\bibinfo {author} {\bibfnamefont {A.~V.}\ \bibnamefont
  {Glushkov}}, \bibinfo {author} {\bibfnamefont {M.~I.}\ \bibnamefont
  {Pravdin}}, \ and\ \bibinfo {author} {\bibfnamefont {A.}~\bibnamefont
  {Saburov}},\ }\href {\doibase 10.1103/PhysRevD.90.012005} {\bibfield
  {journal} {\bibinfo  {journal} {{Phys. Rev. D}}\ }\textbf {\bibinfo {volume}
  {90}},\ \bibinfo {pages} {012005} (\bibinfo {year} {2014})},\ \Eprint
  {http://arxiv.org/abs/1408.6302} {{arXiv}:1408.6302 [astro-ph.HE]}
  \BibitemShut {NoStop}%
\bibitem [{\citenamefont {Sabourov}\ \emph {et~al.}(2017)\citenamefont
  {Sabourov}, \citenamefont {Glushkov}, \citenamefont {Pravdin} \emph
  {et~al.}}]{bib:10}%
  \BibitemOpen
  \bibfield  {author} {\bibinfo {author} {\bibfnamefont {A.~V.}\ \bibnamefont
  {Sabourov}}, \bibinfo {author} {\bibfnamefont {A.~V.}\ \bibnamefont
  {Glushkov}}, \bibinfo {author} {\bibfnamefont {M.~I.}\ \bibnamefont
  {Pravdin}},  \emph {et~al.},\ }\href {https://pos.sissa.it/301/552/}
  {\bibfield  {journal} {\bibinfo  {journal} {{PoS(ICRC2017)552}}\ } (\bibinfo
  {year} {2017})},\ \bibinfo {note} {{(Proc. of the 35th ICRC, Busan,
  Korea)}}\BibitemShut {NoStop}%
\bibitem [{\citenamefont {Abu-Zayyad}\ \emph {et~al.}(2013)\citenamefont
  {Abu-Zayyad}, \citenamefont {Aida}, \citenamefont {Allen} \emph
  {et~al.}}]{bib:11}%
  \BibitemOpen
  \bibfield  {author} {\bibinfo {author} {\bibfnamefont {T.}~\bibnamefont
  {Abu-Zayyad}}, \bibinfo {author} {\bibfnamefont {R.}~\bibnamefont {Aida}},
  \bibinfo {author} {\bibfnamefont {M.}~\bibnamefont {Allen}},  \emph
  {et~al.},\ }\href
  {http://iopscience.iop.org/article/10.1088/2041-8205/768/1/L1/} {\bibfield
  {journal} {\bibinfo  {journal} {{Astrophys. J. Lett.}}\ }\textbf {\bibinfo
  {volume} {768}},\ \bibinfo {pages} {{L1}} (\bibinfo {year}
  {2013})}\BibitemShut {NoStop}%
\bibitem [{\citenamefont {Ivanov}\ \emph {et~al.}(2003)\citenamefont {Ivanov},
  \citenamefont {Krasilnikov},\ and\ \citenamefont {Pravdin}}]{bib:16}%
  \BibitemOpen
  \bibfield  {author} {\bibinfo {author} {\bibfnamefont {A.~A.}\ \bibnamefont
  {Ivanov}}, \bibinfo {author} {\bibfnamefont {A.~D.}\ \bibnamefont
  {Krasilnikov}}, \ and\ \bibinfo {author} {\bibfnamefont {M.~I.}\ \bibnamefont
  {Pravdin}},\ }\href {\doibase 10.1134/1.1648288} {\bibfield  {journal}
  {\bibinfo  {journal} {{JETP. Lett.}}\ }\textbf {\bibinfo {volume} {78}},\
  \bibinfo {pages} {695} (\bibinfo {year} {2003})}\BibitemShut {NoStop}%
\bibitem [{\citenamefont {Vladimirskii}\ \emph {et~al.}(1985)\citenamefont
  {Vladimirskii}, \citenamefont {Galper}, \citenamefont {Luchkov},\ and\
  \citenamefont {Stepanian}}]{bib:12}%
  \BibitemOpen
  \bibfield  {author} {\bibinfo {author} {\bibfnamefont {B.~M.}\ \bibnamefont
  {Vladimirskii}}, \bibinfo {author} {\bibfnamefont {A.~M.}\ \bibnamefont
  {Galper}}, \bibinfo {author} {\bibfnamefont {B.~I.}\ \bibnamefont {Luchkov}},
  \ and\ \bibinfo {author} {\bibfnamefont {A.~A.}\ \bibnamefont {Stepanian}},\
  }\href {\doibase 10.1070/PU1985v028n02ABEH003852} {\bibfield  {journal}
  {\bibinfo  {journal} {{Sov. Phis.~--- Uspekhi}}\ }\textbf {\bibinfo {volume}
  {28}},\ \bibinfo {pages} {153} (\bibinfo {year} {1985})}\BibitemShut
  {NoStop}%
\bibitem [{\citenamefont {Abeysekara}\ \emph {et~al.}(2018)\citenamefont
  {Abeysekara}, \citenamefont {Archer}, \citenamefont {Aune}, \citenamefont
  {Benbow}, \citenamefont {Bird} \emph {et~al.}}]{bib:13}%
  \BibitemOpen
  \bibfield  {author} {\bibinfo {author} {\bibfnamefont {U.}~\bibnamefont
  {Abeysekara}}, \bibinfo {author} {\bibfnamefont {A.}~\bibnamefont {Archer}},
  \bibinfo {author} {\bibfnamefont {T.}~\bibnamefont {Aune}}, \bibinfo {author}
  {\bibfnamefont {W.}~\bibnamefont {Benbow}}, \bibinfo {author} {\bibfnamefont
  {R.}~\bibnamefont {Bird}},  \emph {et~al.},\ }\href {\doibase
  10.3847/1538-4357/aac4a2} {\bibfield  {journal} {\bibinfo  {journal}
  {{Astrophys. J.}}\ }\textbf {\bibinfo {volume} {861}},\ \bibinfo {pages}
  {134} (\bibinfo {year} {2018})},\ \Eprint {http://arxiv.org/abs/1805.05989}
  {{arXiv}:1805.05989 [{astro-ph.HE}]} \BibitemShut {NoStop}%
\bibitem [{\citenamefont {Lloyd-Evans}\ \emph {et~al.}(1983)\citenamefont
  {Lloyd-Evans}, \citenamefont {Coy}, \citenamefont {Lambert} \emph
  {et~al.}}]{bib:14}%
  \BibitemOpen
  \bibfield  {author} {\bibinfo {author} {\bibfnamefont {J.}~\bibnamefont
  {Lloyd-Evans}}, \bibinfo {author} {\bibfnamefont {R.~N.}\ \bibnamefont
  {Coy}}, \bibinfo {author} {\bibfnamefont {A.}~\bibnamefont {Lambert}},  \emph
  {et~al.},\ }\href {\doibase 10.1038/305784a0} {\bibfield  {journal} {\bibinfo
   {journal} {{Nature}}\ }\textbf {\bibinfo {volume} {305}},\ \bibinfo {pages}
  {784} (\bibinfo {year} {1983})}\BibitemShut {NoStop}%
\bibitem [{\citenamefont {Samorsky}\ and\ \citenamefont
  {Stamm}(1983)}]{bib:15}%
  \BibitemOpen
  \bibfield  {author} {\bibinfo {author} {\bibfnamefont {M.}~\bibnamefont
  {Samorsky}}\ and\ \bibinfo {author} {\bibfnamefont {W.}~\bibnamefont
  {Stamm}},\ }\href@noop {} {\bibfield  {journal} {\bibinfo  {journal}
  {{Astrophys. J.}}\ }\textbf {\bibinfo {volume} {263}},\ \bibinfo {pages}
  {{L17}} (\bibinfo {year} {1983})}\BibitemShut {NoStop}%
\end{thebibliography}%

\end{document}